# Determination of Neutrino Mass Hierarchy and $\theta_{13}$ With a Remote Detector of Reactor Antineutrinos

John. G. Learned[1], Stephen T. Dye[1,2], Sandip Pakvasa[1], and Robert C. Svoboda[3,4]

[1]*Department of Physics and Astronomy, University of Hawaii*
[2]*College of Natural Sciences, Hawaii Pacific University*
[3]*Lawrence Livermore National Laboratory*
[4]*University of California at Davis*

We describe a method for determining the hierarchy of the neutrino mass spectrum and $\theta_{13}$ through remote detection of electron antineutrinos from a nuclear reactor. This method utilizing a single, 10-kiloton scintillating liquid detector at a distance of 50-64 kilometers from the reactor complex measures mass-squared differences involving $\nu_3$ with a one (ten) year exposure provided $\sin^2(2\theta_{13})>0.05$ (0.02). Our technique applies the Fourier transform to the event rate as a function of neutrino flight distance over neutrino energy. Sweeping over a relevant range of $\delta m^2$ resolves separate spectral peaks for $\delta m^2_{31}$ and $\delta m^2_{32}$. For normal (inverted) hierarchy $|\delta m^2_{31}|$ is greater (lesser) than $|\delta m^2_{32}|$. This robust determination requires a detector energy resolution of $3.5\%/\sqrt{E}$.

## 1. Introduction

Neutrinos have different masses by virtue of their well established mixing and oscillations [1]. Knowledge of the spectrum of neutrino masses is currently incomplete. We know $\nu_2$ to be more massive than $\nu_1$ ($m_2>m_1$) with $\delta m^2_{21} = (7.9\pm0.7)\times10^{-5}$ eV$^2$ [2]. Although we know $|\delta m^2_{31}|\approx|\delta m^2_{32}| = (2.5\pm0.5)\times10^{-3}$ eV$^2$ [3,4], we do not know if the hierarchy is normal ($m_3>m_2$) or inverted ($m_3<m_1$). The hierarchy can be determined by measuring both $|\delta m^2_{31}|$ and $|\delta m^2_{32}|$ with a precision better than $\delta m^2_{21}/|\delta m^2_{31}|\approx0.03$. For normal (inverted) hierarchy $|\delta m^2_{31}|$ is greater (lesser) than $|\delta m^2_{32}|$. Determination of neutrino mass hierarchy is fundamental to the development of models of particle physics [6] with significant implications for cosmology and astrophysics.

The expression for the survival probability of electron neutrinos involving 3-neutrino mixing is given by [7,8]

$P_{ee}=1-\{\cos^4(\theta_{13})\sin^2(2\theta_{12})\sin^2(\Delta_{21}) + \cos^2(\theta_{12})\sin^2(2\theta_{13})\sin^2(\Delta_{31}) + \sin^2(\theta_{12})\sin^2(2\theta_{13})\sin^2(\Delta_{32})\}$,

where $\theta_{12}$ and $\theta_{13}$ are mixing angles, $\Delta_{ij}=1.27(|\delta m^2_{ji}|L)/E_\nu$ controls the oscillations with $\delta m^2_{ji}\equiv m^2_j-m^2_i$ the neutrino mass-squared difference of $\nu_j$ and $\nu_i$ in eV$^2$, $L$ is the neutrino flight distance in meters, and $E_\nu$ is the neutrino energy in MeV. Three terms, each oscillating with a "frequency" in $L/E$ space specified by $\delta m^2_{ji}$, suppress the survival probability an amount determined by the mixing angles. At present we know $\theta_{13}$ is small [9] and $\theta_{12}$ is large and less than $\pi/4$ [2]. The first term with the lowest "frequency" dominates the suppression. It is responsible for the deficit of solar neutrinos and the conspicuous spectral distortion of reactor antineutrinos [2]. For non-zero $\theta_{13}$ the second term provides greater suppression than the third term. Clearly the ability to measure oscillations influenced by mass-squared differences involving $\nu_3$ requires $\theta_{13}\neq0$. Sensitivity to these oscillations is greatest when $\Delta_{21}=\pi/2$, which provides maximum suppression by the dominant term and thereby the highest signal to noise ratio. For the normal hierarchy of neutrino masses ($m_3>m_2>m_1$) $\Delta_{31}$ is slightly greater than $\Delta_{32}$ giving the second term a slightly higher "frequency" than the third term. Whereas for the inverted hierarchy of neutrino masses ($m_2>m_1>m_3$) $\Delta_{31}$ is slightly smaller than $\Delta_{32}$ giving the second term a slightly lower "frequency" than the third term. It is thus possible to determine neutrino mass hierarchy by resolving the small (~3%) difference in the "frequency" of the second and third terms.

There is discussion in the literature of various methods to determine neutrino mass hierarchy using reactor antineutrinos. These explore the potential for measuring distortions of the energy spectrum due to non-zero $\theta_{13}$ [9,10]. We describe below a unique and robust method.

## 2. Precision Measurement of Mass-Squared Differences Involving $\nu_3$

Neutrino oscillation experiments using reactor antineutrinos are well established. These traditionally involve electron antineutrino disappearance as described by equation (1). Using the standard event rate spectrum we generate data samples in a scintillating liquid detector with an energy resolution of 3.5%/$\sqrt{E}$. The neutrino event spectrum peaks at about 3.6 MeV. This suggests an optimum baseline distance of $L=\pi(3.6\ \text{MeV})/\{2.54(7.9\pm0.7)\times10^{-5}\ \text{eV}^2\}=56\pm7$ km for measuring oscillations involving $\nu_3$. The effect of neutrino mixing on the reactor event spectrum at a distance of 50 km is exhibited by a broad modulation of $\Delta_{21}$ producing a local minimum of event rate at neutrino energy just above 3 MeV. Superposed, for non-zero $\theta_{13}$, is the narrow modulation of $\Delta_{31}$ (assuming normal hierarchy). There is a broadening of the $\Delta_{21}$ and $\Delta_{31}$ modulations with increasing neutrino energy. Plotting the event rate as a function of neutrino flight distance divided by neutrino energy ($L/E$) makes the modulations uniform as we show in Figure 1 for a 1000 kT-y exposure of a detector fixed at 50 km.

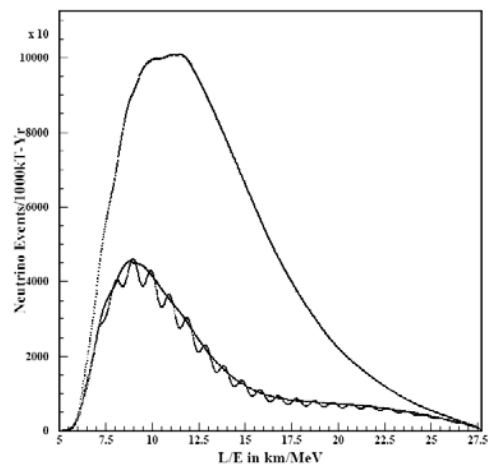

FIG. 1: Event rate versus $L/E$ in units of km/MeV for: no oscillations (top curve), oscillations with $\theta_{13}=0$ (lower smooth curve), and oscillations with $\sin^2(2\theta_{13})=0.1$.

The new approach we describe in this paper utilizes the power of transform methods to extract the signal due to non-zero $\theta_{13}$. We show in Figure 2 the Fourier transform of the data expected for an exposure of 1000 kT-y at a distance of 50 km from an 8 GW$_t$ reactor complex. The transform samples 1000 bins in $L/E$ space, while sweeping over values of $\delta m^2$. At small $\delta m^2$ the spectrum is dominated by the broad $\Delta_{21}$ modulation. It is not possible using this technique to measure the $\delta m^2$ value associated with this feature because only about one cycle of the $\Delta_{21}$ modulation is present in reactor neutrinos at a distance of 50 km. The prominent peak in the spectrum is due to the many cycles of $\Delta_{31}$ modulation, allowing measurement of $\theta_{13}$. This peak measures $\delta m^2_{31}$ with a precision of about 1% for $\sin^2(2\theta_{13})=0.1$.

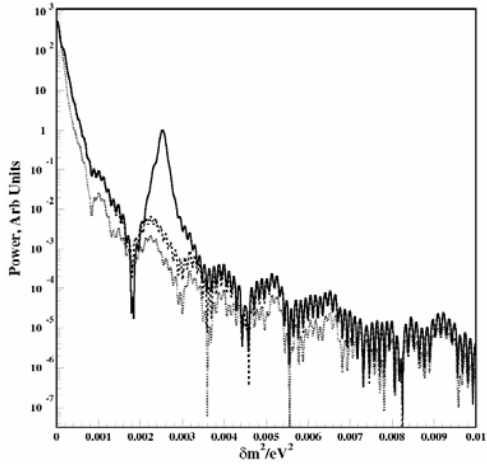

FIG. 2: Fourier power spectrum with modulation in units of eV$^2$ and power in arbitrary units on the logarithmic scale. The peak due to $\Delta_{31}$ with $\sin^2(2\theta_{13})=0.1$ is prominent.

Using this technique it is possible to determine the neutrino mass hierarchy by resolving the small shoulder displaced by $\delta m^2_{21}$ from the main peak. The shoulder with a power reduced by about a factor of 6 is at smaller $\delta m^2$ for normal hierarchy and at larger $\delta m^2$ for inverted hierarchy. We show in Figure 3 just the top of the peak for the two possible hierarchies, where $\delta m^2_{31}$ and $\delta m^2_{21}$ are fixed at experimental values given above.

In order to assess the quantitative ability of an experiment to discriminate between normal and inverted hierarchy, we have written a simulation program which generates and analyzes data sets from an idealized 8.5x10$^{32}$ free proton detector and 8 GW$_{th}$ reactor complex. We have varied the range, $\sin^2(2\theta_{13})$, and exposure time typically for 1000 simulated experiments at each set of parameters.

We have not at this stage included detector specific background sources such as those due to cosmic ray muons traversing the detector, radio impurities, geophysical neutrinos, or neutrinos from other (more distant) reactors. The cosmic ray induced background depends upon depth of water or rock overburden, so must be assessed for the individual proposed location. We know, however that this is of no concern at depths greater than 3 kmwe, though lesser depths may be acceptable. Other reactors will make a small contribution, if sites are chosen on the basis of not having significant additional flux (though to a certain extent these can be included in the analysis). In general we do not expect background to compromise the proposed method, since the added neutrinos start at random distances relative to the detector, so make no coherent contribution to the Fourier transform on $L/E$ at the "frequency" of interest. One may think of such background, if uniformly distributed in *L/E* as simply contributing to the zeroth term in the transform, the total rate. Of course, the more random events in a finite sample, the more background across the $\delta m^2$ spectrum. In any event, at this stage we neglect background, reserving the study for more specific applications.

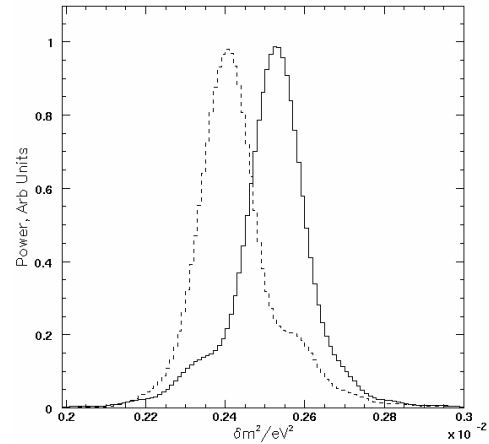

FIG. 3: Neutrino mass hierarchy (normal=solid; inverted=dashed) is determined by the position of the small shoulder on the main peak.

We have studied several algorithms for determining the mass hierarchy, noting that the periodicity ($\delta m^2$), if evident, is measured to 0.1% precision. In practice this is limited by systematic uncertainties in terms of interpretation as a particular mass difference, probably the energy scale uncertainty (of order 1%). However, in the data set the peak is known to whatever we fit it to, and we can analyze the data employing that knowledge. Hence, knowing the primary peak ($\delta m^2_{31}$), we need to determine if the secondary peak is above or below. While we do not know $\delta m^2_{21}$ exceedingly accurately, we know $\sigma_{12}/\delta m^2_{31}$ very well (to about $3\times10^{-3}$). This is to be compared to the spread of about 3% between the two peaks. Hence we can construct a measure examining how well the data fit each hierarchy hypothesis. For presentation here, we use a "matched filter" approach, which one can think of as the Fourier transform of the correlation function, producing a numerical value for each hypothesis.

In Figure 4 we show in a scatter plot the distribution of "experimental" results at distances of 30, 40, 50, and 60 km with normal and inverted hierarchy. Each experiment yields two numbers, the output of the matched filter, which we plot on the x and y axes. One sees that there is very nice separation along the diagonal. Hence we construct a new variable by projecting the distributions onto a 45 degree line. This is illustrated in Figure 5 in four panels. The data fits well to a Gaussian distribution. Separation is quite good (>95%) over the entire range examined, from 30-75 km, but falls off below 40 km and above 65 km.

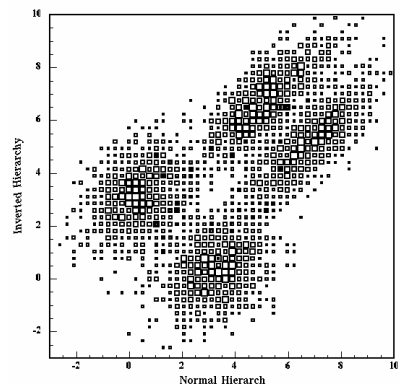

FIG 4: Distance dependent scatter plots for hierarchy test. The plots on the lower right are sets of 1000 experiments at 30 and 50 km with normal hierarchy. Those on the upper left are with inverted hierarchy.

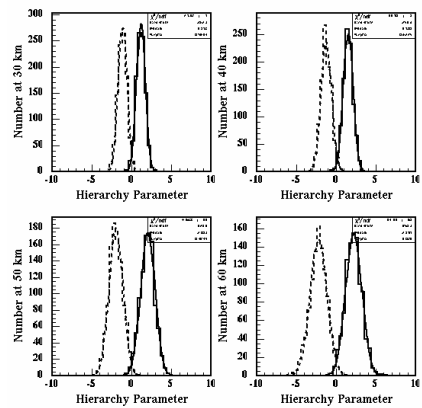

FIG 5: Hierarchy parameter distributions for 30, 40, 50, and 60 km. Solid histograms are with normal hierarchy, dashed with inverted. Distributions fit well to a Gaussian.

Next we examine the sensitivity of the hierarchy determination to $\sin^2(2\theta_{13})$. In Figure 6 we present scatter plots of hierarchy tests for 1000 experiments at each of $\sin^2(2\theta_{13}) = 0.04$, 0.12 and 0.20, all at 50 km range. One sees that the distributions are well separated at $\sin^2(2\theta_{13})$ values more than about 0.04 (in one year). The values of the hierarchy parameter are plotted in the same projection as above for the distance study, in Figure 7. It thus appears as though such an experiment can probe the hierarchy down to $\sin^2(2\theta_{13})$ values of 0.02 with an exposure of 100 kT-y (with the caveats about site specific background).

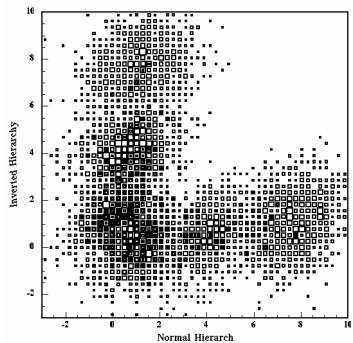

FIG 6: $\sin^2(2\theta_{13})$ dependent scatter plots for hierarchy test using matched filter output. Horizontal plots are sets of 1000 experiments at $\sin^2(2\theta_{13})$ = 0.04, 0.12, and 0.20 with normal hierarchy. Vertical plots are with inverted hierarchy. Note the greater separation with larger $\theta_{13}$.

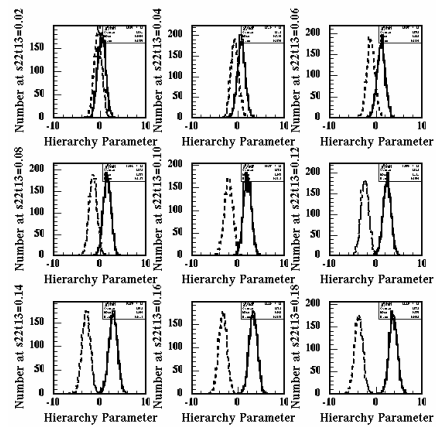

FIG 7: Hierarchy parameter distributions for 1000 experiments each with $\sin^2(2\theta_{13})$ values of 0.02, 0.04, 0.06, 0.08, 0.10, 0.12, 0.14, 0.16, and 0.18. Solid histograms are with normal hierarchy, dashed with inverted.

## 3. Conclusions

We demonstrate a robust method using a single remote detector of reactor antineutrinos that measures $\theta_{13}$ by employing a Fourier transform, and determines neutrino mass hierarchy by resolving mass-squared differences involving $\nu_3$. This determination is provided with an exposure of 10 (100) kT-y and $\sin^2(2\theta_{13})>0.05$ (0.02). This method does not depend on precise measuring or modeling of the reactor flux spectrum nor observation of matter effects.

**Note Added:**

The hierarchy determination is sensitive to the actual values of $\delta m^2_{31}$ and $\delta m^2_{32}$. This is explored is a subsequent paper (M. Batygov et al., *in preparation*). After completion of this work, a related study also employing Fourier transform techniques was reported, which supports the results presented here (L. Zhan et al., arXiv: 0807.3203).